\documentclass{article}

\usepackage[english]{babel}

 \usepackage[a4paper,top=2cm,bottom=2cm,left=3cm,right=3cm,marginparwidth=1.75cm]{geometry}

\usepackage{amsmath}
\usepackage{bm}
\usepackage{authblk}
\usepackage{graphicx}
\usepackage[colorlinks=true, allcolors=blue]{hyperref}
\usepackage{soul}
\usepackage{xcolor}
\usepackage{tabularx}
\usepackage{threeparttable}	
\usepackage{threeparttablex}
\usepackage{array}
\usepackage{makecell}
\usepackage{siunitx}
\usepackage{subfigure}
\usepackage{subcaption}
\usepackage{multirow}
\usepackage{multicol}
\usepackage{makecell}
\usepackage{upgreek}
\usepackage{lscape}
\usepackage{longtable}

\usepackage{lineno}

\title{Final assessment of radioactive impurities in the JUNO detector}
\author[37]{Thomas Adam}
\author[14]{Fengpeng An}
\author[64]{Costas Andreopoulos}
\author[47]{Giuseppe Andronico}
\author[59]{Nikolay Anfimov}
\author[49]{Vito Antonelli}
\author[59]{Tatiana Antoshkina}
\author[37]{Jo\~{a}o Pedro Athayde Marcondes de Andr\'{e}}
\author[35]{Didier Auguste}
\author[59]{Nikita Balashov}

% ...
\affil[14]{Sun Yat-Sen University, Guangzhou, China}
% ...
\affil[35]{IJCLab, Universit\'{e} Paris-Saclay, CNRS/IN2P3, 91405 Orsay, France}
% ...
\affil[37]{IPHC, Universit\'{e} de Strasbourg, CNRS/IN2P3, F-67037 Strasbourg, France}
% ...
\affil[47]{INFN Catania and Dipartimento di Fisica e Astronomia dell Universit\`{a} di Catania, Catania, Italy}
% ...
\affil[49]{INFN Sezione di Milano and Dipartimento di Fisica dell Universit\`{a} di Milano, Milano, Italy}
% ...
\affil[59]{Joint Institute for Nuclear Research, Dubna, Russia}
% ...
\affil[64]{The University of Liverpool, Department of Physics, Oliver Lodge Laboratory, Oxford Str., Liverpool L69 7ZE, UK, United Kingdom}
% ...

\begin{document}
\maketitle

\begin{abstract}
The Jiangmen Underground Neutrino Observatory (JUNO) collaboration has completed the construction of the 20,000-ton liquid scintillator detector and the associated muon veto detector system. To meet the physics objectives, the materials used in the detector must exhibit low radioactive contamination. The single-event rate in the fiducial volume (R $<$ 17.2~m) of the scintillator is required to be approximately 7~Hz for energies above 0.7~MeV, resulting in an accidental coincidence background of about 1 event per day for reactor neutrino physics analyses. Since the beginning of the construction phase, we have screened the natural radioactivity content of  thousands of materials, to select those that meet the design background budget.  The radioactive impurity concentrations of the materials ultimately used in the JUNO detector are summarized in this paper. The construction of the entire detector and the subsequent filling of the liquid scintillator were completed in August 2025. From the initial data, the total count rate of natural radioactivity within the detector’s fiducial volume has met the requirements and is sufficient to support the reactor antineutrino analysis.
\end{abstract}

\section{Introduction}
\label{sec1}

The primary physics objective of the Jiangmen Underground Neutrino Observatory (JUNO) is to determine the neutrino mass ordering and measure neutrino oscillation parameters with high precision by detecting reactor antineutrinos~\cite{JUNO-YB,JUNO:2021vlw,juno-nature2026}. Additionally, JUNO can study neutrinos from various astrophysical sources, such as the Sun, core-collapse supernovae, and the Earth. To achieve these physics goals, the intrinsic background of the detector must be sufficiently low. This background is dominated by the natural radioactivity of detector materials and by cosmic-ray-induced backgrounds. The JUNO detector is constructed at a depth of about 650~m, which reduces the cosmic ray component to a sufficiently low level. However, controlling the natural radioactive content of detector materials remains a critical aspect throughout the entire process of detector design, production, and installation, requiring meticulous quality control based on dedicated radioactivity measurements.

\begin{figure}[ht]
    \centering
    \includegraphics[width=\textwidth]{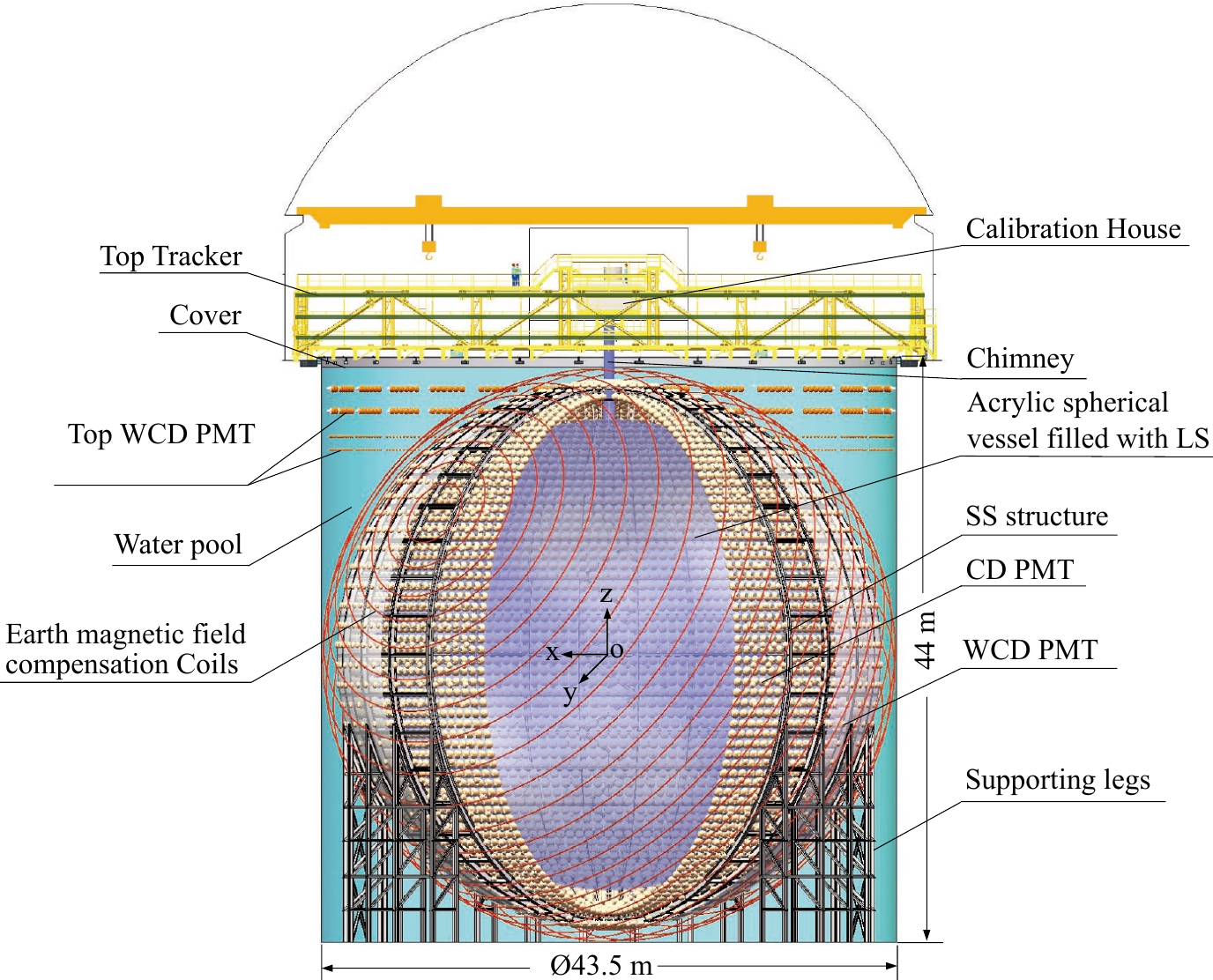}
    \caption{Conceptual diagram of the JUNO detector.}
    \label{fig:detector}
\end{figure}

JUNO is the world's largest liquid scintillator (LS) detector to date: a schematic diagram of its design is shown in Figure\,\ref{fig:detector} and details can be found in Ref.~\cite{JUNO:2021vlw}. Twenty thousand tons of LS are contained within an acrylic sphere with an inner diameter of 35.4~m and a wall thickness of 0.12~m. The 600-ton acrylic vessel is constructed on site by assembling 263 acrylic panels layer by layer. The entire acrylic sphere is supported by an outer stainless steel (SS) structure with a diameter of 41.1~m, connected via SS support rods. The outer surface of the sphere features 590~acrylic nodes for attaching these SS support rods. The central detector (CD) consists of the acrylic vessel and the SS structure. The space outside the acrylic sphere is filled with 40,000 tons of ultra-pure water, which shields against backgrounds from the surrounding rock and serves as a water Cherenkov detector (WCD).

To capture the faint light signals from neutrino interactions, 17,596 20-inch photomultiplier tubes (PMTs) and 25,587 3-inch PMTs are installed on the inner side of the SS truss. Additionally, 2,399 20-inch PMTs are installed on the outer side of the SS truss to observe cosmic rays~\cite{JUNO2026InitialPerformance}. To minimize interference from long cables, the readout electronics and high-voltage modules for the PMTs are located underwater, directly adjacent to the PMTs, requiring extremely high waterproofing standards.

Magnetic shielding coils installed outside the SS truss are used to mitigate the effects of the Earth's magnetic field on the PMTs. A plastic scintillator tracker installed at the detector top assists the water Cherenkov detector in tagging cosmic rays. Since the LS inside the acrylic sphere and the water outside have different densities, the liquid level inside the sphere must be higher than that outside. To accommodate this requirement, a 9-m-high chimney is designed above the acrylic sphere, and connects it to the calibration room at the top. This chimney serves two purposes: it allows for the overflow of LS due to temperature-induced volume changes and it facilitates the deployment of calibration sources into the detector for energy response measurement.

The energy calibration is obtained by a redundant system of multiple sources (both radioactive and laser-based ones) and multidimensional scan systems~\cite{JUNO-Calibration}.
For a one-dimensional scan, the Automatic Calibration Unit (ACU) can deploy sources  along the central axis of the detector.
Off-axis calibration positions are obtained through a Cable Loop System (CLS) that can be moved on a vertical half-plane by adjusting the lengths of two connection cables. During the CLS calibration process, the source position can be verified using four high-resolution Charge-Coupled Device (CCD) cameras (6576×4384) mounted on the equator. These cameras locate an active light source suspended beneath the calibration source, thereby determining the source's position.
A Guide Tube Calibration System (GTCS) surrounds the outside of the acrylic sphere and runs in a longitudinal loop  to calibrate nonuniformity
of the energy response at the detector boundary. A source is pulled into the Teflon GTCS with servomotors
to reach the desired locations. For both the CLS and the GTCS systems an array of eight customized low radioactivity Ultrasonic Sensor System (USS) receivers allows reconstructing
source positions based on signals emitted from transmitters on the source attachment fixture.

The spontaneous decay of naturally occurring radioactive nuclides in detector materials emits particles such as $\alpha$, $\beta$, and $\gamma$ rays, which can enter the LS and contribute to the background of neutrino events. Charged particles such as $\alpha$ and $\beta$ emitted by materials outside the LS are strongly attenuated and rarely reach the LS, whereas neutral $\gamma$ rays can penetrate into the LS and contribute to the background. As $\gamma$ rays propagate through matter, their energies are progressively attenuated; therefore, stricter radioactivity control is required for materials located closer to the LS than for those farther away. Based on the JUNO software framework, we simulated the natural radioactivity on the outer surface of the acrylic sphere, PMTs, and the SS truss~\cite{JUNO:2021kxb}. The radioactivity contributions from the PMTs and SS truss are approximately three orders of magnitude lower than the acrylic outer surface due to the presence of about 3~m of water shielding. The requirements on the concentration of naturally occurring radioactive nuclides inside the LS are particularly stringent, since all decay products can deposit energy in the LS and contribute directly to the background. Based on a combination of industry-wide production capacity surveys and simulation studies, the design specifications for radioactive impurity levels in various detector materials have been defined, and the resulting event rates in the LS are summarized in Table~\ref{tab:bkgBudget}~\cite{JUNO:2021kxb}. The table shows the target impurity concentrations for $^{238}$U, $^{232}$Th, $^{40}$K, $^{210}$Pb and $^{60}$Co. In the simulation, it is assumed that the uranium and thorium decay chains are in secular equilibrium. In the first row, \emph{LS-reactor} refers to the minimum radiopurity requirements for the LS to accomplish the primary physics objectives of JUNO using reactor antineutrinos. The simulated event rates reported in the last two columns of the table correspond, respectively, to the entire detector volume (DV) and to a fiducial volume (FV) with radius R~$<$~17.2~m, for energies above 0.7~MeV. In the table, rows for the same material are subcategorized into separate entries. This accounts for contributions from materials deployed at various locations and different types of PMTs. Even for the same material type, its background contribution can vary depending on installation position. The background levels of products manufactured by different PMT manufacturers vary significantly. Therefore, these are listed in distinct rows. In the "Other" row, we have grouped several minor contributors. This includes calibration components and cables installed on the inner surface of the acrylic sphere (0.12~Hz), the water from the Cherenkov detector (0.06~Hz), and the rock wall (0.13~Hz). Notably, the water in WCD is required to have a U/Th content better than 10$^{-14}$~g/g, a $^{222}$Rn concentration better than 10~mBq/m$^3$, and a $^{226}$Ra concentration better than 1~mBq/m$^3$. For the filling water and the water extraction system, the required radiopurity must be 1–2 orders of magnitude better than that of the WCD water. Specifically, for the filling water, which only contacts the surface of the LS, the radiopurity is required to be one order better than the WCD water. In contrast, for the water extraction system used to purify the LS, the water is fully mixed with the LS; therefore, its radiopurity must be two orders of magnitude better than the WCD water.

To achieve the objectives outlined in Table~\ref{tab:bkgBudget}, we meticulously screened thousands of samples during the selection and production of detector raw materials. Finally, we chose those with the lowest levels of natural radioactivity for the construction of JUNO.

\begin{table}[ht]
\begin{center}
\footnotesize
\renewcommand\arraystretch{1.3}
\begin{tabular}{c|c|c|c|c|c|c|c|c}
\hline
   \multirow{2}{*}{Material} & \multirow{2}{*}{Mass} & \multicolumn{5}{c|}{Target impurity concentration} & \multicolumn{2}{c}{Singles} \\
   \cline{3-9}
& & $^{238}$U & $^{232}$Th & $^{40}$K & $^{210}$Pb & $^{60}$Co  & DV & FV \\
& [t] & [10$^{-9}$~g/g] & [10$^{-9}$~g/g] & [10$^{-9}$~g/g] & [10$^{-9}$~g/g]
& [mBq/kg] & [Hz] & [Hz]  \\
\hline
\hline
LS-reactor & 20000 & 10$^{-6}$ &  10$^{-6}$ & 10$^{-7}$ &  10$^{-13}$
& & 2.5 & 2.2  \\ \hline
Acrylic &  610 & 10$^{-3}$ & 10$^{-3}$ & 10$^{-3}$ & & & 8.4 & 0.4  \\ \hline
\multirow{2}{*}{SS structure} &  1000 & 1 & 3 & 0.2 & & 20 & \multirow{2}{*}{15.9}  & \multirow{2}{*}{1.1} \\
   & 65 & 0.2 & 0.6 &  0.02 && 1.5  && \\ \hline
\multirow{3}{*}{PMT glass} & 33.5 & 400  & 400 &  40 && & \multirow{3}{*}{26.2} & \multirow{3}{*}{2.8} \\
& 100.5 & 200  & 120 &  4 && &&\\
& 2.6 & 400  & 400 &  200 && &&\\
\hline
\multirow{2}{*}{PMT readout} & 125 &68 & 194 & 5 &&16& \multirow{2}{*}{3.4} & \multirow{2}{*}{0.4} \\
& 16.3 & 93 & 243 & 12 &&14&& \\
\hline
Other && &&&& & 2.5 & 0.3 \\ \hline
\hline
\multicolumn{7}{c|}{Sum} & 59 &  7.2  \\
\hline
\end{tabular}
  \caption{Final background budget for the main materials used in the JUNO detector with reconstructed energy larger than 0.7~MeV, taken from Ref.~\cite{JUNO:2021kxb}. The two rows of the SS structure correspond, respectively, to the SS shell positioned farther from the LS, and the SS joints and supporting bars located closer to it. The three rows labeled PMT glass correspond sequentially to 20-inch dynode-PMTs, 20-inch MCP-PMTs, and 3-inch PMTs. The two rows under PMT readout correspond to the readout electronics for 20-inch and 3-inch PMTs, respectively. The expected count rates are given both in the full detector volume (R = 17.7~m) and in the default fiducial volume (R = 17.2~m).  }
\label{tab:bkgBudget}
\end{center}
\end{table}

 The structure of this paper is organized as follows. Section~\ref{sec2} provides a brief overview of the screening facilities and instrumentation  within the collaboration used to quantify the radiopurity of the detector components. Section~\ref{sec3} summarizes the measurement results for all raw materials used in the JUNO detector. Section~\ref{sec4} presents  results from first detector data and compares them with the design expectations. Finally, Section~\ref{sec5} provides a summary of this research.

\section{Screening equipment resources}
\label{sec2}
The methods commonly used to measure the natural radioactivity (U, Th, K) content in materials mainly include three techniques: High Purity Germanium (HPGe) gamma spectrometry, Neutron Activation Analysis (NAA), and Inductively Coupled Plasma Mass Spectrometry (ICP-MS). Each of these detection methods has its own advantages and limitations, and their main performance characteristics are compared in Table~\ref{tab:method}. For a few acrylic samples, the U/Th surface contamination has also been measured by Laser Ablation ICP-MS (LA-ICP-MS).
To ensure a Radon level in the water as low as possible, several facilities for Radon emanation and Radon permeability measurements have been utilized.
A detailed description of the screening facilities used for JUNO samples is provided in Ref.~\cite{JUNO:2021kxb}, to which we refer for further details. In practical measurements, the selection of an appropriate analytical method is made based on a comprehensive consideration of the design specifications and the strengths and limitations of the available techniques. For particularly critical components, multiple methods may be applied.

\begin{table*}[htbp]
\renewcommand\arraystretch{1.3}
\begin{minipage}[c]{\textwidth}
      \vspace{0.5cm}
  \resizebox{\textwidth}{!}{
	\begin{tabular}{c|c|c|c}
        \hline
        & HPGe & NAA & ICP-MS \\
        \hline
        \hline
        Nuclei & $^{238}$U, $^{226}$Ra, $^{228}$Ra,$^{228}$Th, $^{40}$K, $^{60}$Co & $^{238}$U, $^{232}$Th, $^{40}$K, $^{85}$Kr & $^{238}$U, $^{232}$Th \\ \hline
        Destructive testing & No & Yes/No & Yes \\ \hline
        Decay chain equilibrium check & Yes & No & No \\ \hline
        Treatment & No & Simple & Complex  \\ \hline
        Sample mass [kg] & 0.1-20 & 0.001-0.05 & 0.001-0.01  \\ \hline
        Sample mass with enrichment [kg] & - & 0.05-2 & 0.1-10  \\ \hline
        Sensitivity [g/g] & 10$^{-8}$-10$^{-11}$ & 10$^{-12}$-10$^{-16}$ & 10$^{-12}$-10$^{-17}$ \\ \hline
        Time [day] & 1-30 & 10-30 & 2-5 \\ \hline
        \hline
    \end{tabular}}
    \caption{The table presents a comparison of the advantages and disadvantages of the three detection methods. The range of sensitivity and testing time depends on the type of sample.}
	\label{tab:method}
\end{minipage}
\end{table*}

Compared to the instrumentation and sensitivity reported in Ref.~\cite{JUNO:2021kxb}, our quality inspection capabilities have been further enhanced. For LS screening, novel methods have been developed for both ICP-MS and NAA. These methods involve extracting U/Th from the LS using nitric acid, followed by concentration of the acid solution. For ICP-MS, after 3-5 days of experimental processing, the measurement sensitivity for U/Th can reach levels of 10$^{-16}$ to 10$^{-17}$~g/g~\cite{Li:2024gqe}. This technique has been extensively applied in the quality assurance/quality control (QA/QC) processes during the production and filling of the LS for JUNO. For NAA, following the pre-concentration step and neutron irradiation, a radiochemical separation combined with $\beta-\gamma$ coincidence spectroscopy was implemented, allowing sensitivities for U/Th in the range of 10$^{-15}$ to 10$^{-16}$~g/g~\cite{naa-radioch-u-th}. In the case of K, no pre-concentration step is required; only a radiochemical separation is performed after neutron irradiation of the LS, followed by gamma spectroscopy. This method allows sensitivities at the level of 10$^{-16}$~g/g~\cite{naa-K}.

For radiopurity measurements of acrylic samples, we have developed a rapid and sensitive method to quantify $^{238}$U and $^{232}$Th at below $<$10$^{-12}$~g/g within 2–3 days, based on microwave ashing coupled with ICP-MS analysis~\cite{Li:2023rae}. Compared to the previous approach~\cite{Cao:2020zyr}, this method is significantly simpler, safer, and more efficient. It has been extensively implemented in the QA/QC procedures (described in Section~\ref{3.1}) during the production and surface treatment of acrylic components for the JUNO experiment. Cross-check measurements with NAA, using the method described in \cite{naa-2025}, have been performed periodically.

We have significantly enhanced the sensitivity of our instrumentation for measuring radon levels in both gas and water. The achieved sensitivities reach 0.26~$\upmu$Bq/m$^3$ for radon measurements in nitrogen~\cite{Ling:2024ohp} and 0.6~mBq/m$^3$ for radon measurements in water~\cite{Wang:2025mvc}. In addition, by using manganese wire to concentrate radium in water and coupling it with radon detection devices, we have developed a method capable of measuring radium concentrations in water down to 6~$\upmu$Bq/m$^3$~\cite{Li:2024lcy}. This method has been employed for quality control of ultrapure water in JUNO.

\section{Screening results}
\label{sec3}
In this section, the quality inspection results for all detector materials that meet the JUNO requirements are systematically summarized by detector subsystem, following the categories listed in Table~\ref{tab:bkgBudget}.
These results are the outcome of several years of screening activities, during which only materials with sufficiently low contaminant concentrations were retained.
Because different analytical techniques probe different radionuclides within the natural $^{238}$U and $^{232}$Th decay chains, the quality control results are organized accordingly. Measurements of the parent nuclides, obtained using ICP-MS and NAA, are presented  separately from those assuming decay chain equilibrium, which are derived from HPGe spectrometry. When a sample is annotated with “(XX batches)”, the reported result corresponds to the average of multiple production batches. Results without this notation are based on a single measurement. For critical materials, multiple batches were screened to ensure consistency.

\subsection{Screening by ICP-MS and NAA}
\label{3.1}
These two techniques were used for materials whose radiopurity is of critical importance, requiring the screening of $^{238}$U and $^{232}$Th concentration levels well below 10$^{-12}$~g/g. This applies to all materials very close or in direct contact with the LS, including the LS itself, as listed in Table~\ref{tab:icpms-naa}. For ICP-MS screening of water and LS, sample enrichment is required, and the processed sample mass typically ranges from 0.1 to 10 kg, depending on the target sensitivity. In contrast, other material samples analyzed by ICP-MS generally require only a few grams and do not undergo an enrichment step. For NAA, sample masses of the order of 10--50~g were irradiated directly in the case of organic matrices, while LS samples required pre-concentration, with processed masses up to 2~kg.

\begin{figure}[ht]
    \centering
    \includegraphics[width=\textwidth]{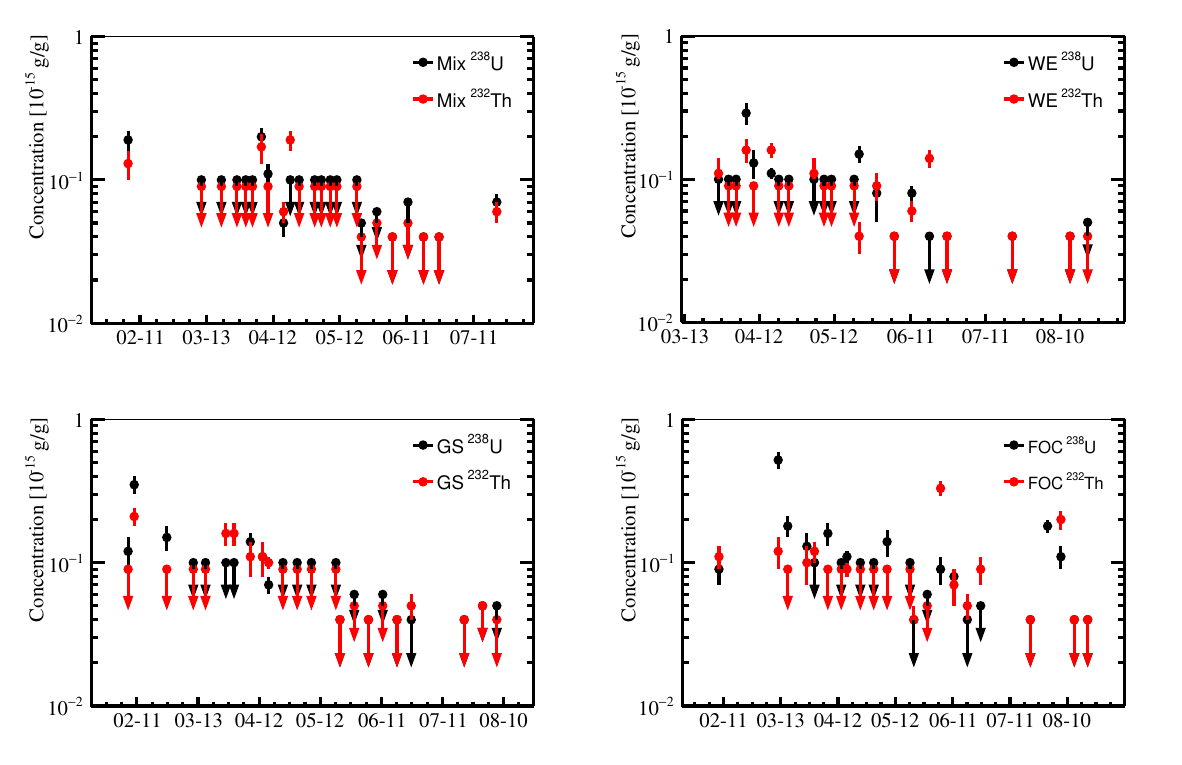}
    \caption{Evolution of ICP-MS screening results on U/Th concentration in LS during the filling of the JUNO detector in the first half of 2025. Top left: "MIX" represents the mixing system. Top right: "WE" represents the water extraction system. Bottom left: "GS" represents the gas stripping system. Bottom right: "FOC" represents the filling and overflow control system.}
    \label{fig:LSmonitor}
\end{figure}

The final JUNO LS mixture consists of four components present at different concentrations. The base solvent is linear alkylbenzene (LAB), which is doped with 2.5~g/L of 2,5-diphenyloxazole (PPO) as primary fluor, 3 mg/L of 1,4-bis(2-methylstyryl)benzene (bis-MSB) as wavelength shifter, and 42.7 mg/L of butylated hydroxytoluene (BHT) as antioxidant. Table~\ref{tab:icpms-naa} reports the results of the radiopurity screening for the individual components.
The LS mixture was prepared and purified directly at the JUNO site~\cite{JUNO:2021vlw}. The raw LAB produced by SINOPEC first undergoes alumina (Al$_2$O$_3$) filtration to reduce optical impurities, and is subsequently processed through a distillation plant~\cite{LS-plants-2024} to remove heavy contaminants. After that, the mixing system produces the LS cocktail, which is further purified underground through water-extraction to remove heavy metals, and through a gas-stripping plant~\cite{LS-plants-2024} to eliminate volatile impurities. The purified LS is finally stored in the Filling and Overflow Control (FOC) system before being transferred into the acrylic vessel. During the LS production for JUNO, regular samples were taken after each purification step to monitor and validate the radiopurity. The results screened by ICP-MS~\cite{Li:2024gqe} are summarized in Figure~\ref{fig:LSmonitor}. The screening results from the distillation system during the commissioning period were consistently below the ICP-MS detection limit. Therefore, samples up to this stage were analyzed only at the beginning of production, while subsequent samples were taken from later stages (mixing, water extraction, gas stripping, and FOC systems) to optimize the use of measurement resources. For the water extraction system, an initial cleanliness issue was identified at the start of production. Consequently, this system was used only after its cleanliness had been confirmed by ICP-MS, and it became operational on March 25, 2025. Most of the screening results in the figure were below the ICP-MS detection limit. In June and July 2025, some of the FOC system screening points yielded measured contaminant concentrations rather than upper limits, while the gas stripping samples only provided upper limits. This discrepancy is tentatively attributed to possible sampling contamination, though it has not been conclusively verified. Table~\ref{tab:icpms-naa} summarizes the average LS screening results—obtained after all purification steps defined by the JUNO protocol~\cite{JUNO:2021vlw} and prior to detector filling—from the FOC system.

Water is another crucial material for the JUNO detector. In addition to being the main component of the water Cherenkov detector, it was used in the water-extraction purification step~\cite{LS-plants-2024} of the LS as well as in the \textit{water-exchange} filling procedure \cite{JUNO2026InitialPerformance} adopted to fill the JUNO detector. After filling the acrylic sphere with ultra pure water (UPW), we collected surface water samples from the top chimney. We have implemented extensive measures to ensure detector cleanliness~\cite{Zhao:2025udt}, and this measurement serves as an indicator of both the internal environment and the acrylic surface cleanliness. The results for the different samples are reported in Table~\ref{tab:icpms-naa}, all of which meet the filling water requirement of U/Th$<$10$^{-15}$~g/g. This indicates that both the internal environment of the acrylic and its surface remain very clean after installation. The $^{222}$Rn and $^{226}$Ra activities in water were measured using the methods described in Refs.~\cite{Wang:2025mvc,Li:2024lcy}. For the water extraction system, the measured $^{222}$Rn activity in UPW was below 1~mBq/m$^3$, while for the water Cherenkov detector it was below 10~mBq/m$^3$. The $^{226}$Ra activity in UPW was below 10~$\upmu$Bq/m$^3$ for both the water Cherenkov detector and the water extraction system.
\begin{figure}[ht]
    \centering
    \includegraphics[width=0.5\textwidth]{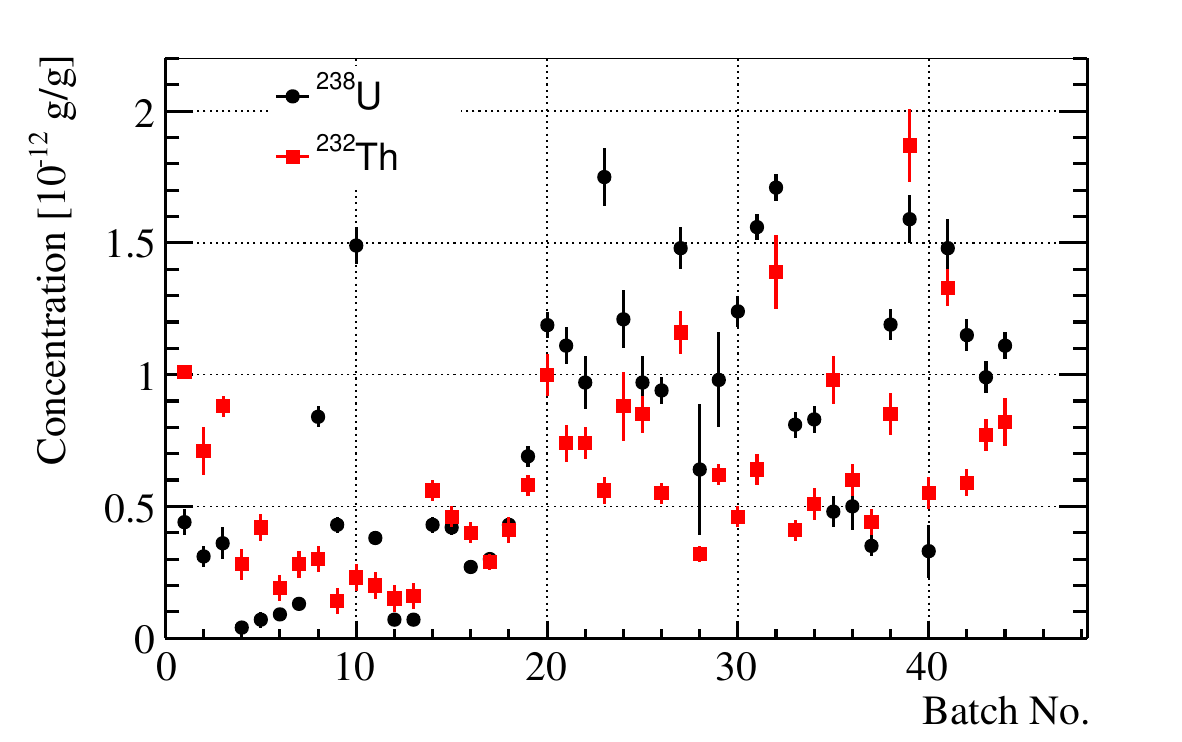}
    \caption{Evolution of ICP-MS screening results on U/Th concentration in acrylic during the production of the panels for the acrylic vessel. }
    \label{fig:acrylicMonitor}
\end{figure}

The acrylic vessel, as anticipated in Section~\ref{sec1}, was constructed on site by bonding 263 acrylic panels layer by layer. All batches of acrylic panels were screened by ICP-MS to validate their radiopurity, with the results shown in Figure~\ref{fig:acrylicMonitor} and the average values summarized in Table~\ref{tab:icpms-naa}. After production, the panels were stored at the manufacturer’s facility for a period of time, protected by a polyethylene (PE) film on the surface. LA-ICP-MS measurements revealed that the radiopurity in the top 50~$\upmu$m of the surface can be 2–3 orders of magnitude higher than that of the bulk material. Based on this, it was decided to remove at least 100~$\upmu$m from the surface prior to shipment to the JUNO site.
Following a detailed study of potential background contamination throughout the surface treatment process, the final procedure was implemented in the JUNO acrylic panel production~\cite{Li:2023rae}. In Table~\ref{tab:icpms-naa}, \emph{acrylic surface} refers to surface scrapings from a given thickness of the acrylic panels, measured to assess the residual radioactive contamination after surface polishing. After careful surface treatment, each panel was covered with a water-soluble paper film to protect it from dust and radon daughters implantation until detector filling; moreover, each bonding line was sealed with a special epoxy to prevent chemical reactions with the LS over the years. Acrylic covers were also manufactured to protect the 20-inch PMT glass from chain explosions-- a phenomenon where the explosion of one PMT generates a shockwave in water that triggers the explosion of adjacent PMTs, resulting in a chain reaction.

\subsection{Screening by HPGe}
For HPGe detectors, only gamma-ray lines from the decay daughters of the $^{238}$U and $^{232}$Th chains—starting from $^{226}$Ra and $^{228}$Ra onward—can be measured. For sufficiently high concentrations of $^{238}$U, sometimes also $\gamma$-ray lines from $^{234}$Th and $^{234m}$Pa may be observed. However, in our background simulations, we assume secular equilibrium throughout the entire decay chains. The gamma spectrometers used for material screening in JUNO are HPGe detectors protected by passive and/or active shieldings and spread in several underground laboratories around the world (China JinPing underground Laboratory (CJPL) in China , Laboratoire Souterrain de Modane (LSM) in France, Laboratori Nazionali del Gran Sasso (LNGS) in Italy) or sea-level laboratories (IHEP in China, LP2i Bordeaux in France, UNIMIB in Italy). The sample mass ranges from 0.1 to 15~kg, depending on the geometry and density of the material.

A system-by-system overview of the HPGe quality screening for JUNO materials is presented in Tables~\ref{tab:hpge-cd}–\ref{tab:veto}.

Table~\ref{tab:hpge-cd} reports the HPGe screening results for most of the materials used to build the central detector (CD) and LS purification system, namely the LS, the acrylic vessel, the SS truss, and the supporting rods. The LS and FOC measurement refers to the stainless steel used for the LS purification system and the FOC system~\cite{JUNO:2021vlw}, i.e. all pipes and tanks employed for the LS filling and level control in the detector.

Table~\ref{tab:calibration} reports the HPGe screening results of most of the materials used in the calibration system, namely the USS, CLS, GTCS and CCD.

Table~\ref{tab:PMT} reports the HPGe screening results for the main materials used in the PMT system, namely the glass bulb, the potting materials, and the voltage divider components. Of the nearly 20,000 20-inch PMTs deployed in the JUNO detector, approximately 5,000 were dynode-PMTs procured from Hamamatsu, while the remaining 15,000 were MCP-PMTs supplied by NNVT. The glass envelopes for the MCP-PMTs were manufactured in China, which provides an opportunity to collaborate closely with the supplier to optimize the production line. The production of the PMT bulbs required extensive screening of low-background raw materials and careful control of possible secondary contamination introduced during manufacturing, originating from the environment, tools, or contact materials. To mitigate these effects, a dedicated background control program was implemented at the manufacturer site, including on-site supervision by members of the collaboration and the optimization of cleanliness procedures throughout the production chain~\cite{Zhang:2017ocm}. These measures significantly reduced the U, Th, and K contents of the MCP-PMT glass reaching levels comparable to the best values reported for large photomultiplier tubes. During mass production, representative batches of bulbs were periodically screened by HPGe to verify the radiopurity, and the average results are summarized in the table. Since all PMTs and their associated electronics are operated underwater, the waterproof encapsulation at the PMT rear end must meet extremely high standards. The measurement data for these potting materials are compiled in the "Potting" section.

Table~\ref{tab:electronics} reports the HPGe screening results for most of the materials used in the electronics system, including the housing, heat dissipation structure, high-voltage module, Global Control Unit (GCU) board, front-end board for 3-inch PMTs, as well as the cables and waterproof flexible conduits that run from the PMT back-end to the dry electronics racks in the experimental hall.

Table~\ref{tab:veto} reports the HPGe screening results for most of the materials used in the veto system, including the earth magnetic field compensation coils, the high-density polyethylene (HDPE) used to prevent the diffusion of radon from rock strata and groundwater, and the truss column base structure fixed at the bottom of the pool.

\begin{landscape}
\begin{table*}[ht]
\renewcommand\arraystretch{1.3}
  \begin{minipage}[c]{1.2\textwidth}
      \vspace{0.5cm}
  \resizebox{1.2\textwidth}{!}{
\begin{tabular}{c|c|c|c|c|c|c|c}
    \hline
    Type & Material & Producer & Mass for JUNO & Technique & $^{238}$U [10$^{-12}$~g/g] & $^{232}$Th [10$^{-12}$~g/g] & $^{40}$K [10$^{-12}$~g/g]  \\
    \hline
    \hline
    \multirow{9}{*}{LS} & LAB (37 batches) & SINOPEC, Nanjing, China & 20,000~t & ICP-MS & (0.6$\pm$0.1)$\times$10$^{-3}$  & (1.0$\pm$0.1)$\times$10$^{-3}$   & - \\
    \cline{2-3}
    & PPO (19 batches) & \multirow{6}{*}{HAISO, Wuhan, China} & 60~t &ICP-MS & 0.09$\pm$0.02  & 0.07$\pm$0.02 &  -\\
    & PPO (2 batches) &  & 60~t &  NAA & $<$0.3 & $<$0.5 & 1.24$\pm$0.03\\
    & BHT (2 batches) &  & 1~t & ICP-MS & 0.14$\pm$0.03  & 0.07$\pm$0.04 & - \\
    & BHT &  & 1~t & NAA & $<$3.0 & $<$3.0  & 3.8$\pm$0.2 \\
    & bis-MSB (3 batches) &  & 72~kg & ICP-MS & 2.9$\pm$0.1& 2.0$\pm$0.1  &- \\
            \cline{2-3}
    & Final LS & \multirow{2}{*}{JUNO} & 20,000~t & ICP-MS & $<$3.4$\times$10$^{-5}$ & $<$3.3$\times$10$^{-5}$ & - \\
    & Final LS &  & 20,000~t & NAA & $<$6.5$\times$10$^{-4}$  & $<$1.8$\times$10$^{-3}$  & $<$2$\times$10$^{-4}$  \\ \hline
    \multirow{4}{*}{Water} & For water-extraction & \multirow{4}{*}{JUNO} & 6,700~t & ICP-MS & $<$2.6$\times$10$^{-4}$  & $<$1.0$\times$10$^{-5}$ &\\
    & For detector filling &  & 23,200~t & ICP-MS & (4$\pm$1)$\times$10$^{-4}$ & $<$4$\times$10$^{-4}$  &\\
    & For water Cherenkov detector & & 40,000~t & ICP-MS & (5$\pm$1)$\times$10$^{-4}$ & $<$4$\times$10$^{-4}$  & \\
    & Filled CD water surface &  & - & ICP-MS & (1.05$\pm$0.06)$\times$10$^{-3}$ & (3.7$\pm$0.7)$\times$10$^{-4}$ & \\
        \hline
    \multirow{11}{*}{\makecell{Organic \\ materials}} & Acrylic panels (44 batches) & Donchamp, Taixing, China & 600~t & ICP-MS & 0.62$\pm$0.05  & 0.72$\pm$0.05  & -\\
    & Acrylic panels (13 batches) & Donchamp, Taixing, China & 600~t & NAA & $<$0.2 & $<$0.1  & 0.1-0.4 \\
    & Acrylic surface ($\sim$10 $\upmu$m) & -& - & ICP-MS & 15.2$\pm$0.7 & 24.3$\pm$0.7 &- \\
    & Acrylic subsurface ($\sim$65-75 $\upmu$m) & -& - & LA-ICP-MS & 3.9$\pm$0.6 & 6.8$\pm$0.8 &- \\
    & PMT acrylic cover & Huashuaite, Jiaxing, China & 110~t & ICP-MS & 3.1$\pm$0.3 & 10$\pm$1 & - \\
    & Teflon in acrylic node & BMC, Kunshan, China & 300~kg & NAA & 16$\pm$2 & 82$\pm$5 & 283$\pm$11 \\
    & Epoxy (type 8010) for acrylic bonding line & DODI, Shanghai, China & 3~kg & ICP-MS & $<$5  & $<$7  & - \\
    & Water-soluble glue on film & Elleair, Japan & - & ICP-MS & 450$\pm$20 & 115$\pm$6 & - \\  \cline{2-8}
    & Calibration CLS anchors, PTFE & Sanxin Inc., China & 6.3~kg & NAA & $<$1.2 & 10$\pm$2 & 1.50$\pm$0.05 \\
    & Calibration wear-resisting blocks, PTFE & Sanxin Inc., China & 6.9~kg & NAA & $<$1.2 & 10$\pm$2 & 1.50$\pm$0.05 \\
    & Calibration GTCS Tube, PTFE & Jianwei, China & 20~kg &   NAA & $<$1.65 & $<$3 & 470$\pm$10 \\
    \hline
    \hline
    \end{tabular}}
    \caption{This table lists the JUNO samples analyzed using the destructive testing methods ICP-MS and NAA. The upper limits for most materials listed in this table are quoted at 95\% C.L., whereas those for calibration components are given at 90\% C.L.}
      \label{tab:icpms-naa}
\end{minipage}
\end{table*}

\begin{table*}
\renewcommand\arraystretch{1.3}
  \begin{minipage}[c]{1.2\textwidth}
      \vspace{0.5cm}
  \resizebox{1.2\textwidth}{!}{
	\begin{tabular}{c|c|c|c|c|c|c|c|c}
              \hline
    CD and LS & Material & Producer & Mass for JUNO & Screening &  $^{226}$Ra [Bq/kg] & $^{228}$Ra [Bq/kg] & $^{40}$K [Bq/kg]  & $^{60}$Co [mBq/kg]  \\
    \hline
    \hline
    \multirow{2}{*}{LS} & PPO & HAISO, Wuhan, China & 60~t & UNIMIB & $<$0.2 & $<$0.03 & $<$0.3 & $<$0.01\\
    & Al$_2$O$_3$ & SHAN LV YI FENG, Zibo, China & 300~t & IHEP & 0.26$\pm$0.03 & $<$0.22 & $<$0.70 & - \\
    \hline
    \multirow{3}{*}{\makecell{Acrylic \\ node}} & Bulk, SS304 (2 batches) &  \multirow{2}{*}{TISCO, Taiyuan, China}  & 24~t  & CJPL & $<$4$\times$10$^{-3}$ &  (2.8$\pm$0.7)$\times$10$^{-3}$ & $<$0.05 & 0.7$\pm$0.3 \\
    & Bolt, SS304 &  && CJPL &  (7$\pm$5)$\times$10$^{-3}$  & (8$\pm$2)$\times$10$^{-3}$ & 0.05$\pm$0.04 & 2$\pm$1 \\ \cline{2-3}
    & Solder $\Phi$1.2, SS308 & Golden Bridge, Tianjin, China & 4.5~t & IHEP &  0.39$\pm$0.08 &0.15$\pm$0.10 & 2.4$\pm$0.5 & -\\ \hline

    \multirow{2}{*}{Support rod} & Bulk, SS304 (8 batches) & \multirow{2}{*}{TISCO, Taiyuan, China}   & 67~t & CJPL &  $<$4$\times$10$^{-3}$ &  (2$\pm$1)$\times$10$^{-3}$ & 0.05$\pm$0.03 & 2.0$\pm$0.7 \\
     & Bulk, SS304 & & 67~t  & LNGS & (0.3$\pm$0.1)$\times$10$^{-3}$ &(0.8$\pm$0.3)$\times$10$^{-3}$ & $<$2.8$\times$10$^{-3}$ & (0.4$\pm$0.1)$\times$10$^{-3}$ \\
    \hline
    \multirow{6}{*}{SS truss} & Bulk, SS304 (17 batches) & \multirow{5}{*}{TISCO, Taiyuan, China}   & 608~t  & CJPL & (6$\pm$2)$\times$10$^{-3}$ & (3$\pm$1)$\times$10$^{-3}$ & (4$\pm$2)$\times$10$^{-3}$ & 1.4$\pm$0.5 \\
    & Bulk, SS304 &   & 608~t  &  LP2IB & $<$1.1$\times$10$^{-3}$ &  (4.7$\pm$0.7)$\times$10$^{-3}$ & $<$9$\times$10$^{-3}$ & 0.50$\pm$0.15 \\
    & Bolt, SS630, SS &   & 65~t &  CJPL & $<$4$\times$10$^{-3}$ &  (6$\pm$3)$\times$10$^{-3}$ & 0.17$\pm$0.07 & 10$\pm$3 \\
    & Ring gasket, SS &   & & CJPL & $<$4$\times$10$^{-3}$ &  (1.5$\pm$0.7)$\times$10$^{-2}$ & 0.3$\pm$0.1 & 11$\pm$2 \\
    & Spring, SS &  & 10~t & IHEP & $<$0.11 & $<$0.06 & $<$0.35 & - \\  \cline{2-3}
    & Paint, SS & NIMTE, Ningbo, China & & CJPL &  (2.4$\pm$0.7)$\times$10$^{-2}$ &  (1.9$\pm$0.3)$\times$10$^{-2}$ & 0.43$\pm$0.05 & 3$\pm$2 \\ \cline{2-3}
    & Solder $\Phi$3.2, SS308 & Golden Bridge, Tianjin, China & 8.9~t & IHEP &  0.17$\pm$0.07 & 0.21$\pm$0.06 & $<$1.3  &-\\   \hline
    & Bulk, SS304 & \multirow{3}{*}{TISCO, Taiyuan, China} && IHEP & $<$0.12 & $<$0.15 & $<$0.6 & -\\

    Chimney & Bellow, SS304 & & & CJPL & (8$\pm$5)$\times$10$^{-3}$ & $<$5$\times$10$^{-3}$ & 0.04$\pm$0.03 & 12$\pm$2 \\
    & Bolt, SS304 & & 48~kg & CJPL &  (1.3$\pm$0.7)$\times$10$^{-2}$ &  (4$\pm$2)$\times$10$^{-3}$ & 0.09$\pm$0.05 & 11$\pm$2 \\
    \hline
    LS and FOC & Tank and pipes, SS316 (2~batches) & TISCO, Taiyuan, China   & 50~t  & CJPL &  $<$4$\times$10$^{-3}$ &  (6$\pm$2)$\times$10$^{-3}$ & 0.09$\pm$0.05 & 3$\pm$2 \\ \hline
    \multirow{2}{*}{Acrylic Vessel} & Acrylic sample & \multirow{2}{*}{Donchamp, Taixing, China} & 600~t & LP2IB &  (400$\pm$0.1)$\times$10$^{-6}$ &  $<$220$\times$10$^{-6}$ & $<$2.3$\times$10$^{-3}$ & - \\
     & Acrylic sample &  & 600~t & LNGS & $<$45$\times$10$^{-6}$ & $<$62$\times$10$^{-6}$  & $<$0.3$\times$10$^{-3}$ & - \\ \hline
    \multirow{3}{*}{\makecell{Acrylic surface \\ treatment}} & Sanding paper & Anda, Taixing, China & 10$^6$~pieces & IHEP & $<$0.5 & $<$0.4 &  $<$1.9 & - \\
    & PE film & Baojiali, Nantong, China & 10$^5$~m$^2$ & IHEP & $<$0.23 & $<$0.15 &  $<$0.9 & - \\
    & Protection paper film & Daio Paper Corporation, Japan & 10$^4$~m$^2$ & IHEP & $<$0.5 & 0.6$\pm$0.2 &  $<$1.6 & - \\
    \hline
    \hline
    \end{tabular}}
    \caption{This table lists the CD and LS--related samples analyzed using the non-destructive HPGe testing methods. The upper limits reported in this table are given at 95\% C.L., except for the LNGS measurements, that are quoted at 68\% C.L.}
      \label{tab:hpge-cd}
\end{minipage}
\end{table*}

\begin{table*}
\renewcommand\arraystretch{1.3}
  \begin{minipage}[c]{1.2\textwidth}
      \vspace{0.5cm}
  \resizebox{1.2\textwidth}{!}{
	\begin{tabular}{c|c|c|c|c|c|c|c|c}
    \hline
    Calibration & Material & Producer & Mass for JUNO & Screening &  $^{226}$Ra [Bq/kg] & $^{228}$Ra [Bq/kg] & $^{40}$K [Bq/kg]  & $^{60}$Co [mBq/kg]  \\
    \hline
    \hline
    \multirow{2}{*}{USS} & Cables (PTFE, copper) & Pasterneck, USA & 1.45~kg & CJPL & (3.4$\pm$0.5)$\times$10$^{-3}$ & (1.9$\pm$0.5)$\times$10$^{-3}$ & (1.7$\pm$0.4)$\times$10$^{-2}$ & $<$0.4 \\
    & Receiver (Ni, epoxy, PCB) & Customized, China  & 10 pieces & CJPL & $<$3.3~mBq/piece & $<$2.8~mBq/piece & $<$19~mBq/piece & $<$0.86~mBq/piece  \\ \hline
    CLS & Cables (PTFE, SS) & Fengshuo, China & 0.27~kg & CJPL & $<$0.06 & $<$0.08 & $<$0.5 & $<$38 \\ \hline
    \multirow{2}{*}{GTCS} & Cable (PTFE, SS) & Fengshuo, China & 0.33~kg & CJPL & $<$0.04 & $<$0.04 & $<$0.14 & (2.9$\pm$0.1)$\times$10$^{2}$ \\
    & Sensor (SS, PVC, PBT) & Omron, Japan  & 10 pieces & CJPL & (111$\pm$3)~mBq/piece & (184$\pm$5)~mBq/piece  & (383$\pm$21)~mBq/piece & $<$2.9~mBq/piece  \\ \hline
    \multirow{3}{*}{CCD}& Camera & OPT, Dongguan, China & 4.4~kg & IHEP & 0.7$\pm$0.3 & 3.3$\pm$0.4 & 12$\pm$2 & - \\
    & Cable & Fanya, Zhongshan, China & 0.01~t & IHEP & $<$0.36 & $<$0.27 & $<$1.4 & -\\
    & Fibre-optical & HongXin, Dongguan, China & 0.4~t & IHEP & $<$0.3 & $<$0.17 & $<$1.0 &- \\
    \hline
    \hline
     \end{tabular}}
    \caption{This table lists the JUNO calibration related samples analyzed using the non-destructive testing methods HPGe. The upper limits reported in this table are given at 90\% C.L.}
      \label{tab:calibration}
\end{minipage}
\end{table*}

\begin{table*}[ht]
\renewcommand\arraystretch{1.3}
  \begin{minipage}[c]{1.2\textwidth}
      \vspace{0.5cm}
  \resizebox{1.2\textwidth}{!}{
	\begin{tabular}{c|c|c|c|c|c|c|c|c}
    \hline
    PMT & Material & Producer & Mass for JUNO & Screening &  $^{226}$Ra [Bq/kg] & $^{228}$Ra [Bq/kg] & $^{40}$K [Bq/kg]  & $^{60}$Co [mBq/kg]  \\
    \hline
    \hline
    \multirow{8}{*}{Bulb} & Module, SS304 (3~batches) & \multirow{2}{*}{JISCO, Gansu, China} & 300~t  & CJPL &  $<$0.01 & $<$2$\times$10$^{-3}$ & $<$0.03 & $<$1.5 \\
    & 20-inch PMT cover, SS304 (3~batches) & & 150~t  & CJPL & $<$0.01 & $<$5$\times$10$^{-3}$ & $<$0.03 & $<$3 \\  \cline{2-3}

    & 20-inch MCP-PMT (10 batches) & Huida, Yancheng, China & 90~t  & IHEP & 1.9$\pm$0.2 & 0.6$\pm$0.2 & $<$1.4 & - \\ \cline{2-3}
    & 20-inch dynode-PMT (6 batches) & \multirow{3}{*}{Hamamatsu, Japan} & 30~t & IHEP & 5.3$\pm$0.2 &  1.4$\pm$0.3 & 1.7$\pm$0.2 & - \\
    & 20-inch dynode-PMT &  & 30~t & LP2IB & 6.5$\pm$0.2 & 2.3$\pm$0.1 & 1.9$\pm$0.3&- \\
    & 20-inch dynode-PMT & & 30~t & UNIMIB & 5.4$\pm$0.2 &  1.9$\pm$0.1 & 1.7$\pm$0.3&- \\ \cline{2-3}
    & 3-inch PMT (8 batches) & \multirow{2}{*}{Zhanchuang, Hainan, China} & 2.6~t  & IHEP & 2.2$\pm$0.2 &  1.7$\pm$0.2 & 29$\pm$2 & - \\
    & 3-inch PMT &  & 2.6~t  & LP2IB & 2.7$\pm$0.2 &  2.8$\pm$0.2 & 32$\pm$2 & - \\ \hline

    \multirow{7}{*}{Potting} & 20-inch PMT shell, SS304 (2~batches) & JISCO, Gansu, China & 10~t  & CJPL & $<$6$\times$10$^{-3}$ & (5$\pm$3)$\times$10$^{-3}$ & 0.06$\pm$0.02 & 0.9$\pm$0.5 \\
    & 3-inch PMT shell, ABS & Kemei, Ningbo, China & 6.4~t & IHEP & $<$0.32 & $<$0.22 & 8.2$\pm$0.9 & - \\
    & Putyl tape (type HM36) & AECC BIAM, Beijing, China & 5.7+2.5~t & IHEP &  2.2$\pm$0.2 & 0.4$\pm$0.1 & $<$1.5 & -\\
    & 20-inch PMT Epoxy (type 8101R) & DODI, Shanghai, China & 2.8~t & IHEP & 0.30$\pm$0.04 & 0.08$\pm$0.03 & $<$1 &- \\
    & 3-inch PMT Epoxy (type 4412) & Buffle, Beijing, China & 0.5~t & IHEP & $<$0.2 & 0.39$\pm$0.06 & $<$0.6 &- \\
    & Polyurethane (type 6127) & Buffle, Beijing, China & 15+2~t & IHEP & $<$0.29 & $<$0.21 & $<$1.2 & - \\
    & Shrinkable tube & WOER, Shenzhen, China & 2.5+0.5~t & IHEP & $<$0.27 & $<$0.18 & $<$0.9 & - \\ \hline

    \multirow{11}{*}{Divider} & 20-inch PMT PCB &- & - & IHEP & 15$\pm$1 &  19$\pm$2 & 17$\pm$5 & - \\
    & 20-inch PMT capacitors & Vishay, USA &-& IHEP & 131$\pm$5 & 18.8$\pm$0.9 & 3$\pm$1 & - \\
    & 20-inch PMT resistors & Vishay, USA  & - & IHEP & 2.0$\pm$0.2 &  2.1$\pm$0.2 & 10$\pm$1 & - \\
    & Divider for MCP-PMT & TJcentre, Tianjin, China & 0.44~t & IHEP & 34$\pm$3 & 15$\pm$3 & $<$22 & - \\
    & Divider for dynode-PMT & TJcentre, Tianjin, China & 0.16~t & IHEP & 36$\pm$4 &  19$\pm$3 & 21$\pm$11 & - \\ \cline{2-9}
    & 3-inch PMT PCB &- & 0.11~t & IHEP & 35$\pm$2 &  38$\pm$2 & 52$\pm$3 & -\\
    & 3-inch PMT chip resistors & Viking, Taiwan, China &-& IHEP &  1.4$\pm$0.2 &  2.3$\pm$0.2 & 2.0$\pm$0.8 & - \\
    & 3-inch PMT PCB+resistors & - & 0.11~t & LP2IB & 31$\pm$3 &  31$\pm$2 & 29$\pm$4 & -\\
    & 3-inch PMT capacitors & Murata, Japan & 0.04~t & IHEP &  13.6$\pm$0.6 &  5.6$\pm$0.5 & 4$\pm$1 & - \\
    & 3-inch PMT capacitors & Murata, Japan & 0.04~t & LP2IB &  12.2$\pm$2.2 &  5.8$\pm$1.3 & $<$5.5 & - \\

    & Divider for 3-inch PMT  & Juyingdianlu, Shenzhen, China & 0.25~t & IHEP &  29$\pm$1 &  34$\pm$2 & 42$\pm$3 & - \\
    \hline
    \hline
     \end{tabular}}
    \caption{This table lists the JUNO PMT-related samples analyzed using the non-destructive testing methods HPGe. The upper limits reported in this table are given at 95\% C.L.}
      \label{tab:PMT}
\end{minipage}
\end{table*}

\begin{table*}[ht]
\renewcommand\arraystretch{1.3}
  \begin{minipage}[c]{1.2\textwidth}
      \vspace{0.5cm}
  \resizebox{1.2\textwidth}{!}{
	\begin{tabular}{c|c|c|c|c|c|c|c|c}
    \hline
    Electronics & Material & Producer & Mass for JUNO & Screening &  $^{226}$Ra [Bq/kg] & $^{228}$Ra [Bq/kg] & $^{40}$K [Bq/kg]  & $^{60}$Co [mBq/kg]  \\
    \hline
    \hline
    \multirow{6}{*}{\makecell{Common \\ materials \\ for all PMTs}} & SS shell & GL, Kunshan, China & 78+6.2~t & IHEP & $<$0.11 & $<$0.07 & 0.3$\pm$0.2 & - \\
    & Shrinkable tube & WOER, Shenzhen, China& 0.6~t & IHEP & $<$0.17 & $<$0.14 & $<$0.5 & - \\ \cline{2-3}
    & Thermal gel & GLPOLY, Shenzhen, China & 3.5+0.06~t & IHEP & 0.83$\pm$0.08 &  0.27$\pm$0.06 & $<$0.7 & - \\
    & Thermal copper & Futelang, Suzhou, China & 15~t & IHEP & $<$0.05 & $<$0.03 & $<$0.2 & - \\
    & Back-end cable & Fanya, Zhongshan, China & 68~t & IHEP & $<$0.18 & $<$0.12 & $<$0.6 & - \\
    & Back-end below & Taihe, Wuhu, China & 72~t & IHEP &$<$0.21 & $<$0.13 & $<$0.8 & - \\ \hline

     \multirow{7}{*}{\makecell{20-inch \\ PMTs}} & HVB PCB & GDM, Shanghai, China & 0.20~t & IHEP & 10$\pm$1 &  18$\pm$2 & 17$\pm$6 & - \\
    & HVB metal box (Al) & Juchuangli, Shenzhen, China & 0.22~t & IHEP & $<$0.17 & $<$0.10 & $<$0.49 & - \\
    & Entire HVB & SCC, Shenzhen, China & 0.9~t & IHEP & 3$\pm$1 & 5$\pm$1 & $<$18 & - \\ \cline{2-3}
    & GCU board & SCC, Shenzhen, China & 1.9~t & IHEP & 8.8$\pm$0.5 & 12.0$\pm$0.7 & 13$\pm$1 & - \\
    & Cooling appendix (copper) & Futelang, Suzhou, China & 0.6~t & IHEP & $<$0.1 & $<$0.07 & $<$0.3 & - \\
    & Front-end cable & Fanya, Zhongshan, China & 1.7~t & IHEP & 0.9$\pm$0.4 & 0.5$\pm$0.3 & $<$8 & - \\
    & Front-end below (SS) & Taihe, Wuhu, China & 4.4~t & IHEP & 0.11$\pm$0.05 & $<$0.10 & 0.4$\pm$0.3& - \\ \hline

     \multirow{5}{*}{\makecell{3-inch \\ PMTs}} & HV splitter board & \multirow{2}{*}{SCC, Shenzhen, China} & 0.4~t & LP2IB & 13.8$\pm$1.1 & 17.1$\pm$1.2 & 16.9$\pm$3.0 & - \\
     & GCU board && 0.07~t & LP2IB & 6.0$\pm$0.3 & 7.1$\pm$0.4 & 46.4$\pm$6.1 & - \\ \cline{2-3}
     & ABC front-end board & FEDD, Agôn Electronics, France & 0.07~t & LP2IB & 6.7$\pm$0.5 & 9.3$\pm$0.5 & 5.3$\pm$1.1 & \\
    & Front-end cable & Axon, Foshan, China & 1.4~t & LP2IB & (5.6$\pm$1.7)$\times$10$^{-3}$ & $<$4.2$\times$10$^{-3}$ & $<$25$\times$10$^{-3}$ & $<$1 \\

    & Connector+Front-end cable & Axon, Foshan, China & 2.8~t & IHEP & 0.8$\pm$0.3 &  1.0$\pm$0.3 & $<$1.8 &  \\
    \hline
    \hline
     \end{tabular}}
    \caption{This table lists the JUNO electronics related samples analyzed using the non-destructive testing methods HPGe. The upper limits reported in this table are given at 95\% C.L.}
      \label{tab:electronics}
\end{minipage}
\end{table*}

\begin{table*}[ht]
\renewcommand\arraystretch{1.3}
  \begin{minipage}[c]{1.2\textwidth}
      \vspace{0.5cm}
      \resizebox{1.2\textwidth}{!}{
	\begin{tabular}{c|c|c|c|c|c|c|c|c}
    \hline
    Veto & Material & Producer & Mass for JUNO & Screening &  $^{226}$Ra [Bq/kg] & $^{228}$Ra [Bq/kg] & $^{40}$K [Bq/kg]  & $^{60}$Co [mBq/kg]  \\
    \hline
    \hline
    & Earth magnetic field coils & Zhenshi, Jiaxing, China & 168~t & IHEP & $<$0.16 & $<$0.09 & $<$0.4 &- \\
    & HDPE & GSE, Thailand  & 7500~m$^2$ & IHEP & $<$0.19 &  0.17$\pm$0.05 & 0.8$\pm$0.5 & - \\
    & Truss column base, SS & TISCO, Taiyuan, China & 2~t & IHEP & $<$0.16 & $<$0.10 & $<$0.46 & - \\
    & Solder & Golden Bridge, Tianjin, China &-& IHEP & $<$0.14 & $<$0.09 & $<$0.4 &- \\
    \hline
    \hline
     \end{tabular}}
    \caption{This table lists the JUNO veto related samples analyzed using the non-destructive testing methods HPGe. The upper limit in this table is at 95\% C.L.}
      \label{tab:veto}
\end{minipage}
\end{table*}

\end{landscape}

\subsection{Screening by Radon facilities }
The Radon belonging to the $^{238}$U chain is a radioactive noble gas able to diffuse through barriers or to emanate from the JUNO materials immersed in the water pool. The JUNO requirement for radon in the water pool is 10~mBq/m$^3$.
To prevent radon from rock and groundwater from diffusing into the ultrapure water, a 5-mm-thick HDPE layer was laid flat against the rock surface. The shielding effectiveness can be evaluated based on radon’s permeability through HDPE (so-called transparency), which is determined by placing a radon source on one side of the HDPE and measuring the radon concentration on the opposite side to calculate the radon penetration ratio. The transparency of a 5 mm HDPE liner has been measured with a dedicated apparatus in two cases: one with air on each side of the HDPE liner and one with water on one side and air on the other side (more realistic case for JUNO). The results are given in Table~\ref{tab:liner}. It is observed that the presence of water enhances the Radon transparency by a factor 5 to 6. Nevertheless, the 5-mm HDPE layer remains effective at blocking radon from the rock side in the case of the water condition.

\begin{table*}[ht]
\renewcommand\arraystretch{1.3}
\centering
  \begin{minipage}[c]{0.9\textwidth}
      \vspace{0.5cm}
        \resizebox{0.9\textwidth}{!}{
	\begin{tabular}{c|c|c|c|c}
 \hline
    Material & Producer & Configuration & Screening & Transparency \\
    \hline
    \hline
    \multirow{2}{*}{5 mm HDPE liner } & \multirow{2}{*}{GSE, Thailand} & air / air (3 samples)  & \multirow{2}{*}{CPPM} & (2.6$\pm$0.4)$\times$10$^{-5}$  \\
                    &   & air / water (2 samples) &  & (14.2$\pm$0.9)$\times$10$^{-5}$  \\
        \hline
        \hline
    \end{tabular}}
    \caption{This table lists the Radon transparency measurements for HDPE liner.}
    \label{tab:liner}
\end{minipage}
\end{table*}

Table~\ref{tab:Radon-emanation} lists the JUNO samples screened by Radon emanation measurements in nitrogen, the bare 20-inch and 3-inch PMTs as well as the HDPE liner. More details of the setup are given in \cite{rn-eman}.

\begin{table*}[ht]
\renewcommand\arraystretch{1.3}
\centering
  \begin{minipage}[c]{\textwidth}
      \vspace{0.5cm}
  \resizebox{\textwidth}{!}{
	\begin{tabular}{c|c|c|c|c}
 \hline
    Material & Producer & Quantity measured / Total & Screening & Radon emanation \\
    \hline
    \hline
    5 mm HDPE liner & GSE, Thailand & 2 m$^2$ / $\sim$6500 m$^2$ & LP2IB & $<$1.8 mBq/m$^2$\\
    20-inch bare PMTs & Huida, Yancheng, China & 2 / 15056 & LP2IB & $<$1.9 mBq/PMT  \\
    20-inch bare PMT & Hamamatsu, Japan & 1 / 4939 & LP2IB & $<$3.4 mBq/PMT  \\
    3-inch bare PMTs & Zhanchuang, Hainan, China & 29 / 25587 & LP2IB & $<$0.13 mBq/PMT \\
        \hline
        \hline
    \end{tabular}}
    \caption{This table lists the Radon emanation measurements. The upper limits in this table are given at 90\% C.L.}
    \label{tab:Radon-emanation}
\end{minipage}
\end{table*}

\subsection{Concentration of \texorpdfstring{$^{14}$C}{14C} in LAB}

The radionuclide $^{14}$C is an important background source in the low-energy region relevant for the solar neutrino analyses. Furthermore, its high decay rate can cause pile-up with other physics events within a single readout window, resulting in degraded energy resolution and thereby substantially impacting reactor antineutrino measurements. For these reasons, the $^{14}$C concentration was measured in several LAB samples, including those produced for the JUNO experiment. The detector used for this purpose is located in the JUNO underground laboratory (650 m.w.e.) and is surrounded by a multi-layer passive shield consisting of acid-cleaned oxygen-free high-conductivity (OFHC) copper, lead, 1~cm of HDPE, and 60~cm of borated polyethylene for neutron attenuation. The entire setup is enclosed in a light-tight, radon-suppressed environment maintained by a continuous purge of high-purity nitrogen ($>$100~L/h). The system employs 0.86~kg of LS contained in a 1-L transparent acrylic vessel (76~mm inner diameter, 2~mm wall thickness), viewed by two low-radioactivity Hamamatsu R1140-20 PMTs. These PMTs exhibit a quantum efficiency of 35\% at 420~nm and a U/Th radiopurity of approximately 1~mBq per PMT. A coincidence trigger--set at a threshold of nearly 1~photoelectron (p.e.) with a 44~ns time window--is used to suppress dark noise and random background events.
The key results obtained with this setup are summarized in Table~\ref{tab:C14}. The systematic uncertainty of 20\% is dominated by the detector configuration and the light yield.

\begin{table}[ht]
\renewcommand\arraystretch{1.3}
\begin{center}
\footnotesize
\renewcommand\arraystretch{1.3}
	\begin{tabular}{c|c}
 \hline
      Sample Identifier & $^{14}$C Concentration [10$^{-17}$~g/g] \\
      \hline
      \hline
        Average of 8 times sampling in 2023 & 2.8 $\pm$ 0.2 (stat.) $\pm$ 0.6 (sys.)\\
        OSIRIS storage tank (2023.11)        & 3.4 $\pm$ 0.2 (stat.) $\pm$ 0.7 (sys.) \\
        JUNO LAB tank (2024.11.20)        & 3.3 $\pm$ 0.4 (stat.) $\pm$ 0.7(sys.) \\
     \hline
     \hline
    \end{tabular}
    \caption{Summary of \textsuperscript{14}C concentration measurements in various LAB samples. }
    \label{tab:C14}
    	\end{center}
\end{table}

\section{First JUNO data}
\label{sec4}
JUNO completed the construction of the detector and the LS filling on August 26, 2025, and officially began data collection~\cite{JUNO2026InitialPerformance}. In this study, we analyzed the event distribution in the LS using data collected from October 1, 2025 to March 18, 2026, starting approximately one month after the end of the filling. The total live time of the dataset is 140~days. At this stage, the radon introduced during the filling operations had decayed away to a nearly constant background level. The vertex and energy reconstruction for all events follows the procedure detailed in Ref.~\cite{JUNO2026InitialPerformance}, with several post-publication refinements incorporated. The resulting position reconstruction bias is constrained to within 10~cm, and the measured energy resolution amounts to 3.4\% at 1~MeV using the $^{68}$Ge calibration source.

Our objective is to extract the event rate originating from natural radioactivity using experimental data, which requires excluding contributions from non-radioactive events as thoroughly as possible. Simulations indicate that radioactive events typically have energies below approximately 5~MeV. In contrast, cosmic muons possess much higher energies. When traversing the detector, they produce large numbers of secondary particles, some of which may deposit energy within the same low-energy range and thus contaminate the radioactive event sample. Therefore, in this analysis, a single event is defined as an event recorded in the CD  after applying the muon veto.

The definition of muon and muon veto strategy is the same as that in Ref.~\cite{juno-nature2026}. Muon candidates were identified as CD triggers with charge Q$>$3$\times$10$^4$~p.e. or water pool (WP) triggers with Q$>$700~p.e. We vetoed the full CD for 5~ms following a muon event. This results in a live time efficiency of 95\%, which is included in the results presented below.

\begin{figure}[ht]
    \centering
    \includegraphics[width=0.44\textwidth]{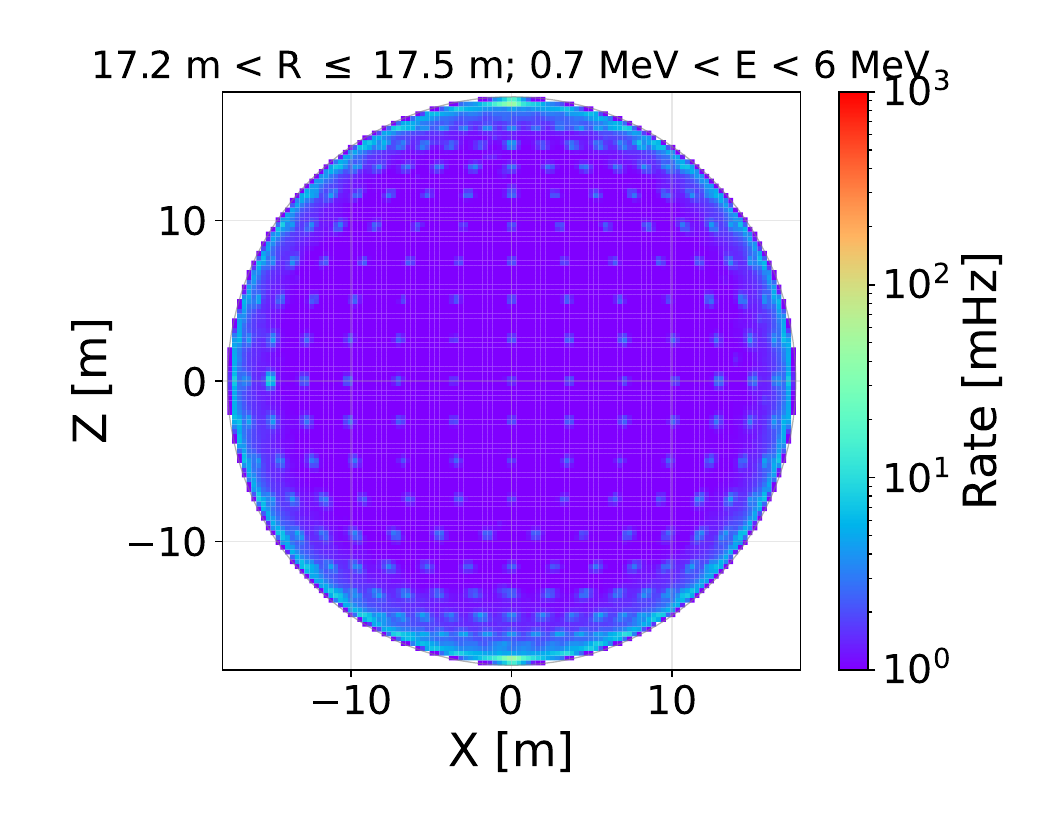}
    \includegraphics[width=0.44\textwidth]{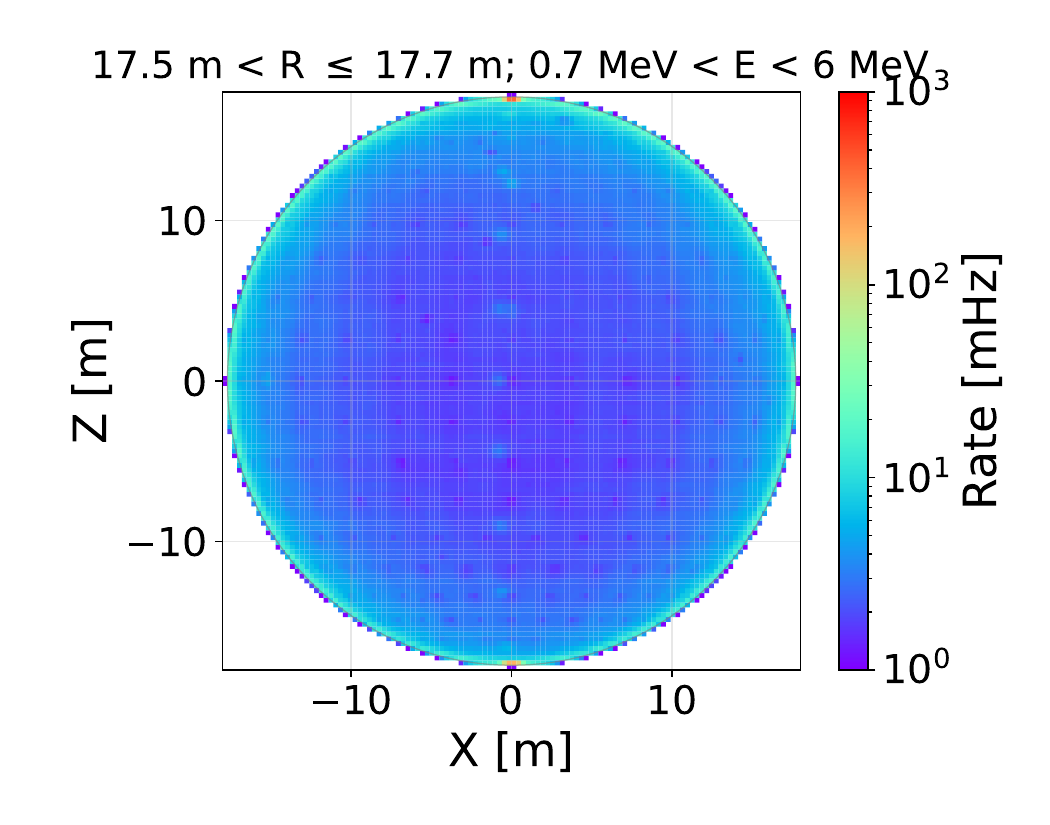}
    \caption{The x–z distributions of the event rate in the LS, in the energy range 0.7–6~MeV, are shown for two different radial cuts: 17.2~m$<$R$<$17.5~m (left) and  17.5~m$<$R$<$17.7~m (right).}
    \label{fig:CDevent}
\end{figure}

Apart from potential reconstruction bias, the top and bottom of the JUNO CD are each equipped with an approximately 1-meter-long acrylic chimney section filled with LS. Given the thinner water shielding in this region and the special optical path of events, some chimney events are reconstructed inside the detector volume (R$<$17.7~m). To address this issue, dedicated variables for identifying chimney events have been developed in the reconstruction algorithm, which can efficiently reject misreconstructed chimney events. A rejection efficiency of 99\% is achieved for the region R$<$17.2~m, with the impact on standard physical events kept below 1\%. A detailed study on chimney event tagging will be thoroughly discussed in the subsequent reconstruction paper. Therefore, the chimney event selection criterion is adopted in the event selection of this study to effectively suppress the interference from chimney events.

The event rate distributions in the external region of the detector for 17.2~m$<$R$<$17.5~m and 17.5~m$<$R$<$17.7~m are represented in the left and right plots of Figure~\ref{fig:CDevent}, respectively. The bright spots in the left figure primarily originate from the background contributions of acrylic nodes and SS support rods on the outer surface of the acrylic vessel. These components are located closer to the detector, and the gamma rays produced by natural radioactive decay can penetrate deeper. In contrast, the events in the right figure are located closer to the edge and mainly originated from the PMTs contribution. Because the PMTs are farther from the LS—with nearly two meters of water shielding—their contribution becomes less significant than that of the acrylic nodes on the outer surface of the acrylic vessel for R$<$17.5~m.

\begin{figure}[ht]
    \centering
    \includegraphics[width=0.5\textwidth]{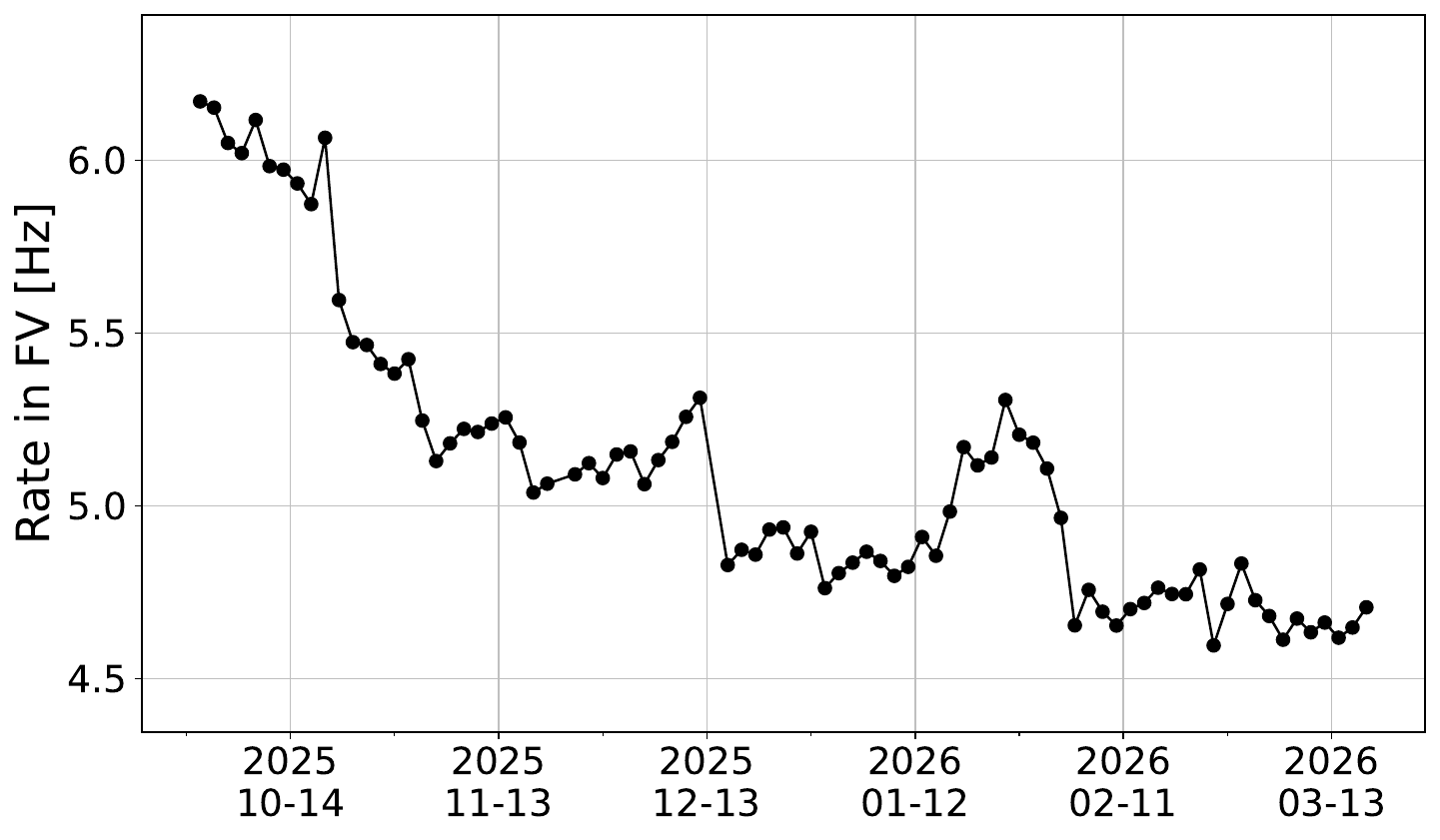}
    \caption{The rate evolution of events with energy larger than 0.7~MeV in the LS FV with R$<$17.2~m as a function of time since October 2025.}
    \label{fig:rate_time}
\end{figure}

Figure~\ref{fig:rate_time} shows the time evolution of the event rate in the LS for the baseline FV (with R$<$17.2~m). A gradual decreasing trend is observed, potentially attributable to the combined effects of residual radon decay and the purification of the circulating veto water. A small rise in event rate was spotted in early December and late January. Its cause is yet unknown, with candidate explanations being water-temperature-induced shifts in liquid scintillator temperature and ensuing convection.

\begin{figure}[ht]
    \centering
    \includegraphics[width=0.44\textwidth]{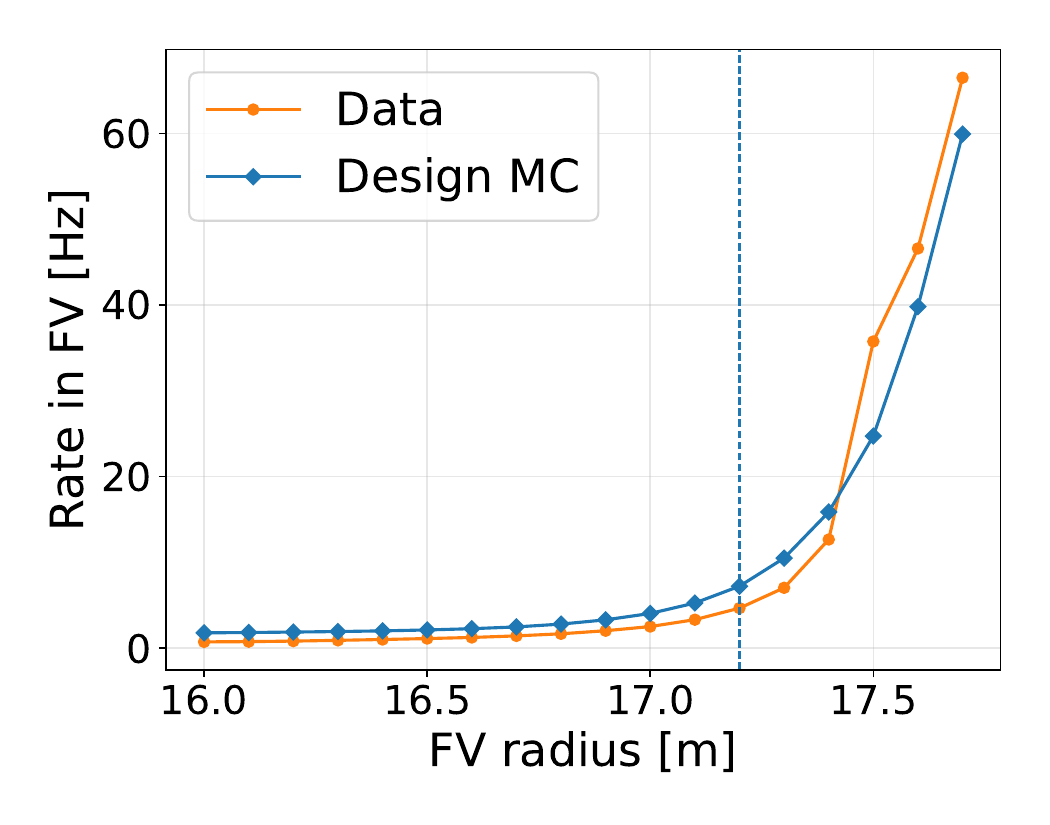}
    \caption{The figure shows the comparison of event rate as a function of the fiducial volume cut between data taken from February 5 to March 18 in 2026 (orange) and MC simulation from Ref.~\cite{JUNO:2021kxb} (blue). The dashed line indicates the baseline selection of R$<$17.2~m. }
    \label{fig:CDvsMC}
\end{figure}

A comparison of the event rate in the LS as a function of position between data (from February 5 to March 18 in 2026) and MC simulation~\cite{JUNO:2021kxb} is shown in Figure~\ref{fig:CDvsMC}. It presents the event rate (for events with E$>$0.7~MeV) under different FV cuts. Each data point represents the total event rate within the volume defined by R$<$FV, where the dashed line indicates the baseline selection of R$<$17.2~m. A detailed numerical comparison between data and simulation is provided in Table~\ref{tab:CDvsMC}. The leading systematic originates from the event reconstruction, whose performance will be further improved in future iterations. The JUNO Collaboration has developed multiple event reconstruction algorithms. Relative to alternative reconstruction implementations, discrepancies in the FV singles rates currently stand at approximately 20-30\%. Note that this is mainly due to systematic differences in the reconstructed positions of events occurring near the acrylic vessel. Elsewhere, all the algorithms agree with each other very well. From the results, it can be observed that the radioactive background from LS and detector materials measured within the baseline FV is better than the design value from MC simulations. This improvement is consistent with the measured internal radiopurity of the LS, which is approximately one order of magnitude below the design baseline requirement of 10$^{-15}$~g/g for U/Th. As indicated in Table~\ref{tab:bkgBudget}, this deficit would contribute $\sim$2 Hz to the LS event rate.

\begin{table}[ht]
\begin{center}
\footnotesize
\renewcommand\arraystretch{1.3}
	\begin{tabular}{c|c|c}
        \hline
        \multicolumn{2}{c|}{Rate [Hz]}  & Fiducial volume \\ \hline
        \multirow{2}{*}{Design} & LS (10$^{-15}$~g/g U/Th)  & 2.2 \\
        & Other detector material  & 5 \\ \hline
        Data & LS + Other detector material   & 4.7 \\ \hline
    \end{tabular}
    \caption{The table compares the event rates above 0.7~MeV for the fiducial volume (R$<$17.2~m), as obtained from design simulations and experimental data taken from February 5 to March 18 in 2026.}\label{tab:CDvsMC}
	\end{center}

\end{table}

\section{Summary}
\label{sec5}
To ensure the detector achieved sufficiently low background levels, JUNO implemented a rigorous screening campaign to characterize the natural radioactivity of materials during both the R$\&$D and construction phases, involving the assay of several thousand samples. This paper details the quality control results for all materials ultimately deployed in the detector. Using data taken after the completion of LS filling for the JUNO detector, we performed a detailed analysis of the event rate. The results were compared with the original MC simulations. It is found that the radial distribution of the event rate agrees well with the simulations. The event rate inside the FV meets the design specifications and satisfies the requirements for reactor neutrino oscillation studies.

\section*{Acknowledgements}
We are grateful for the ongoing cooperation from the China General Nuclear Power Group. This work was supported in part by: the Chinese Academy of Sciences, the National Key R\&D Program of China, the People's Government of Guangdong Province, and the Tsung--Dao Lee Institute of Shanghai Jiao Tong University in China.
We appreciate the contributions from the Institut National de Physique Nucl\'eaire et de Physique des Particules (IN2P3) in France, the Istituto Nazionale di Fisica Nucleare (INFN) in Italy, the Fonds de la Recherche Scientifique (F.R.S.--FNRS) and the Institut Interuniversitaire des Sciences Nucl\'eaires (IISN) in Belgium, the European Structural and Investment Funds, the Czech Ministry of Education, Youth and Sports and the Charles University Research Center in Czech Republic, the Deutsche Forschungsgemeinschaft (DFG), the Helmholtz Association, and the Cluster of Excellence PRISMA+ in Germany, the Joint Institute for Nuclear Research (JINR), the Slovak Research and Development Agency in the Slovak Republic, the MOST and MOE in Taipei, the Program Management Unit for Human Resources \& Institutional Development, Research and Innovation, Chulalongkorn University, and Suranaree University of Technology in Thailand, the Science and Technology Facilities Council (STFC) in the United Kingdom, and the University of California at Irvine and the National Science Foundation (NSF) in the United States.
We also acknowledge the computing resources provided by the Chinese Academy of Sciences, IN2P3, INFN, and JINR, which are essential for data processing and analysis within the JUNO Collaboration.

\bibliographystyle{unsrt}
\bibliography{reference}

\end{document}